\documentclass[rnote]{aa} 

\usepackage{graphicx}
\usepackage{txfonts,ulem}
\begin{document}

\title{Spectroscopic and photometric analysis of the early-type spectroscopic binary HD~161853 in the centre of an H {\sc ii} region}

\titlerunning{Spectroscopic and photometric analysis of the early-type, spectroscopic binary HD~161853}


   \author{
R. Gamen\inst{1} 
\and
C. Putkuri\inst{1}\thanks{While fellow of the Consejo Interuniversitario Nacional of Argentina.}
\and
N. I. Morrell\inst{2} 
\and
R. H. Barb\'a\inst{3} 
\and
J. I. Arias\inst{3} 
\and
J. Ma\'iz Apell\'aniz\inst{4}
\and
N. R. Walborn\inst{5}
\and
A. Sota\inst{6}
\and
E. J. Alfaro\inst{6}
          }

\institute{Instituto de Astrof\'{\i}sica de La Plata (CONICET La Plata - UNLP),
Facultad de Ciencias Astron\'omicas y Geof\'{\i}sicas, Universidad Nacional de La Plata,
Paseo del Bosque S/N, B1900FWA, La Plata, Argentina.\\
              \email{rgamen@fcaglp.unlp.edu.ar}
        \and
Las Campanas Observatory, Carnegie Observatories, Casilla 601, La Serena, Chile.
\and
 Departamento de F\'isica, Universidad de La Serena, Cisternas 1200 Norte, La Serena, Chile.
\and
Centro de Astrobiolog\'ia (CSIC-INTA), ESAC Campus, P.O. Box 78, 28691 Villanueva de la Ca\~nada, Madrid, Spain.
\and
Space Telescope Science Institute\thanks{Operated by AURA, Inc., under NASA contract NAS5-2655}, 3700 San Martin Drive, Baltimore, MD 21218, USA.
\and
Instituto de Astrof\'isica de Andaluc\'ia-CSIC, Glorieta de la Astronom\'ia s/n, 18008, Granada, Spain.
}

   \date{}


  \abstract
  {} 
   {We study the O-type star HD~161853, which has been noted as a probable double-lined spectroscopic 
binary system.}
   {We secured high-resolution spectra of HD~161853 during the past nine years.
We separated the two components in the system and measured their respective radial velocities for
the first time.}
   {We confirm that HD~161853 is an $\sim$1 Ma old binary system consisting of an O8 V star
($M_{\rm A,RV} \geq 22$ M$_\odot$) and a B1--3 V star ($M_{\rm B,RV} \geq 7.2$ M$_\odot$)
at about 1.3 kpc. From the radial velocity curve, we measure an
orbital period $P$ = 2.66765$\pm$0.00001 d and an eccentricity $e$ = 0.121$\pm$0.007. 
Its $V$-band light curve is constant within 0.014 mag and does not display eclipses, from which we impose a maximum orbital inclination $i=54$~deg. HD~161853 is probably associated with an
H {\sc ii} region and a poorly investigated very young open cluster. In addition,
we detect a compact emission region at 50 arcsec to HD~161853 in 22$\mu$m-{\it WISE} and 
24$\mu$m-{\it Spitzer} images, which may be identified as a dust wave piled up by the radiation
pressure of the massive binary system.
}
   {}


   \maketitle
%

\section{Introduction}

\object{HD 161853} (=CPD --31 4999; 
RA~(J2000)=17:49:16.6, Dec~(J2000)=--31:15:18;
$l$=358.4248, $b$=--1.8767; $V_J=7.9$~mag) 
is an O-type star that has been considered in the literature as either a massive or a post-AGB star.
It was first included in the Henry Draper Catalogue \citep{1924hdhc.bookQ....C} and
classified as type B3.  Subsequently, 
\citet{1971AJ.....76..260C,1972MNRAS.158...85C}, who also reported radial velocity variations,
and \citet{1973AJ.....78.1067W} classified it as O7.5 and O8~V((n)), respectively.

New observations of HD~161853 in the mid- and far-infrared led to confusion about 
its true nature. For instance, \citet{1990A&A...233..181R} proposed it to be a planetary nebula candidate 
because its radio emission at 4.8 GHz and $IRAS$ colours were typical of such objects. 
Furthermore, \citet{1993ASPC...45..173P} considered it a post-AGB star rapidly evolving toward a 
young planetary nebula nucleus.
Although \citet{2007A&A...469..799S} disqualified it as a post-AGB star because of its high effective
temperature of more than 35\,000~K, they did not clarify its actual nature, 
and more works continued considering HD~161853 as a low-to-intermediate-mass star 
\citep[e.g. ][]{2011MNRAS.412.1137C}.

HD~161853 is located at the centre of an H {\sc ii} region 
associated with a CO molecular cloud \citep{1982ApJS...49..183B} at 
a kinematic distance of 1.5$\pm$0.2 kpc \citep{2014A&A...569A.125H}.
There has also been some confusion because 
the region was catalogued by \citet{1953ApJ...118..362S,1959ApJS....4..257S}
with different running numbers (Sh~1-17 and Sh~2-15).
In addition, the H {\sc ii} region is not unequivocally identified in the SIMBAD database and has many 
unrelated entries, such as
RCW~134 \citep{1960MNRAS.121..103R}, W~25 \citep{1958BAN....14..215W}, and
G358.464--01.897 \citep{2014ApJS..212....1A}.
The nebula was also identified at H$\alpha$ by \citet{1955MmRAS..67..155G} as Gum~69
(in the SIMBAD database, Gum~69 is used as an alternative identifier of the star HD~161853).

In the field of HD~161853 there are also other objects, suggesting that the area is a young region that deserves further studies.
For example, there are several reported young star candidates \citep{2008AJ....136.2413R}, and
the open cluster Dutra-Bica 51 \citep{2000A&A...359L...9D}. 
\citet{2002A&A...389..871D} determined a distance of 1.3 kpc, an apparent diameter of 2.2 arcmin, 
and an age of only 1 Ma for the cluster.
However, \citet{2004A&A...425..937D} did not find a stellar overdensity in the field that could be related to a cluster.
A supernova remnant, G358.4--01.9, with dimensions 40$\times$36 arcmin, whose centre is 1.5 arcmin south of
HD~161853, was reported by \citet{1988srim.conf..293R}. However,
\citet{1994MNRAS.270..835G} ruled out the existence of such a
remnant because the region presents a 
thermal spectral distribution.
Finally, there is a faint $ROSAT$ source at the position of HD~161853, 
1RXS J174916.5--311509 \citep{2000IAUC.7432R...1V}, also detected by $Chandra$ 
\citep[CXOGBS J174916.6--311518; ][Albacete-Colombo, in prep.]{2014ApJS..210...18J}. 

Multiplicity of HD~161853 has been suspected since the publications of
\citet{1971AJ.....76..260C,1972MNRAS.158...85C}.
More recently, \citet{2012A&A...543A..11M} noted
double He~{\sc i} absorption lines that were related to a binary nature of the source,
and \citet{gosss2} identified them as coming from an O8~V(n)z and B-type components,
but they did not compute any orbit determination.

In this paper, we present the spectral classifications of the
two components of HD~161853 and their 
respective radial velocity orbits. We also analyse the available images of the field and use the 
ASAS $V$-band data to constrain the orbital inclination. 
Finally, we discuss the actual nature of HD~161853.
Preliminary results of this work were shown by \citet{putkuri}.


\section{Observations}

This work is based on observations obtained within the intensive spectroscopic campaign named 
the OWN Survey \citep{2010RMxAC..38...30B}.
We employed the REOSC Cassegrain spectrograph in cross-dispersed mode at the Complejo
Astron\'omico El Leoncito (CASLEO) in Argentina, along with
FEROS at the 2.2m telescope of ESO/La Silla, and the \'echelle spectrograph at the 2.5m du Pont 
telescope of Las Campanas Observatory (LCO), in Chile.
Observations were secured between 2006 May and 2013 August.
The instrumental configuration, observatory, spectral resolving power, 
wavelength coverage, and the number of obtained spectra ($N$) are summarised in Table~\ref{runs}.

At CASLEO and LCO, we obtained a wavelength calibration lamp (Th-Ar) exposure immediately before 
or after each target integration, at the same telescope position. 
The spectra were processed and calibrated using standard IRAF\footnote{
IRAF is distributed by the National Optical Astronomy Observatories, which are operated by
the Association of Universities for Research in Astronomy, Inc., under cooperative
agreement with the National Science Foundation.} 
routines.
For FEROS, we applied the standard reduction pipeline provided by ESO, which 
uses comparison lamp exposures obtained 
at the beginning and end of each observing night.

\begin{table}
\caption{Details of the spectroscopic data for HD~161853.
}
\begin{tabular}{l l c c c}
\hline\hline\noalign{\smallskip}
 Instr. Config.    & Observatory & R     & $\Delta \lambda$ & $N$ \\
                 &           &       & [\AA]    &\\
\noalign{\smallskip}\hline\noalign{\smallskip}
\'Echelle, 2.5-m    & LCO         & 40\,000 &  3450--9850 & 13\\
REOSC, 2.15-m       & CASLEO      & 15\,000 &  3600--6100 & 26\\
FEROS, 2.2-m        & La Silla    & 46\,000 &  3570--9210 &  4\\
\noalign{\smallskip}\hline
\end{tabular}
\label{runs}
\end{table}

\section{Results and discussion}
\subsection{Radial velocities}

Some of our spectra of HD~161853 clearly show two components, as noted by \citet{2012A&A...543A..11M}
and \citet{gosss2}.
Thus, we employed the method for separating composite spectra as explained
in \citet{2006A&A...448..283G}.
After a few iterations, we obtained the individual spectra of HD~161853 A and B and radial velocities 
for the 43 observed epochs.
Some Balmer lines showed issues due to incorrect normalisation of the \'echelle orders
containing broad lines.

We performed the cross-correlation by using the {\sc fxcor} task of IRAF. The wavelength regions
considered were 5008--5024\,\AA\ and 5864--5888\,\AA, which include the He {\sc i} $\lambda5015$ and
$\lambda5875$ absorption lines present in both components. {\sc fxcor} provides errors, but as they are
smaller than the expected instrumental errors, we adopted a conservative error of 2.5 km s$^{-1}$ for
all measurements. Moreover, the interstellar Na~{\sc i} $\lambda\lambda5890,5896$\,\AA\ measured in the spectra
resulted in a mean of $-$10 km s$^{-1}$ with a maximum difference of 2.5 km s$^{-1}$ among them, 
and a standard deviation of 1 km s$^{-1}$. We did not find any systematic differences among
measurements performed with the three spectrographs used in this work.
The Heliocentric Julian Days (HJD) and radial velocities (RV) of the two components are shown in
Table~\ref{vrs}.

\begin{table}
\caption{Radial velocities of the two components in HD~161853.} \label{vrs}
\begin{tabular}{c cc ccc}
\hline\hline\noalign{\smallskip}
HJD &  RV$_A$ & RV$_B$              & Weight & Phase & Obs.\\
&  \multicolumn{2}{c}{[km s$^{-1}$]}&        & $\phi$&     \\
\noalign{\smallskip}\hline\noalign{\smallskip}
2\      453\    867.879 &       25      &       $-36$   &       0.5     &       0.07    &       CAS     \\
2\      453\    875.893 &       33      &       $-84$   &       0.5     &       0.08    &       LCO     \\
2\      454\    222.835 &       63      &       $-192$  &       1.0     &       0.13    &       CAS     \\
2\      454\    246.821 &       58      &       $-174$  &       1.0     &       0.12    &       ESO     \\
2\      454\    251.822 &       $-17$   &       106     &       0.5     &       1.00    &       CAS     \\
2\      454\    252.804 &       84      &       $-236$  &       1.0     &       0.36    &       CAS     \\
2\      454\    253.667 &       $-52$   &       216     &       1.0     &       0.69    &       CAS     \\
2\      454\    257.897 &       94      &       $-276$  &       1.0     &       0.27    &       LCO     \\
2\      454\    258.640 &       5       &       $-16$   &       0.5     &       0.55    &       LCO     \\
2\      454\    258.882 &       $-45$   &       136     &       1.0     &       0.64    &       LCO     \\
2\      454\    286.600 &       32      &        ...    &       1.0     &       0.03    &       CAS     \\
2\      454\    286.753 &       39      &        ...    &       1.0     &       0.09    &       CAS     \\
2\      454\    312.603 &       $-97$   &       296     &       1.0     &       0.78    &       CAS     \\
2\      454\    316.490 &       95      &       $-272$  &       1.0     &       0.24    &       CAS     \\
2\      454\    316.611 &       98      &       $-271$  &       1.0     &       0.28    &       CAS     \\
2\      454\    339.469 &       $-95$   &        ...    &       1.0     &       0.85    &       CAS     \\
2\      454\    339.584 &       $-80$   &       255     &       1.0     &       0.90    &       CAS     \\
2\      454\    340.506 &       99      &       $-267$  &       1.0     &       0.24    &       CAS     \\
2\      454\    340.608 &       104     &       $-268$  &       1.0     &       0.28    &       CAS     \\
2\      454\    608.891 &       $-85$   &       315     &       1.0     &       0.85    &       CAS     \\
2\      454\    643.765 &       $-70$   &       252     &       1.0     &       0.92    &       CAS     \\
2\      454\    659.757 &       $-74$   &       248     &       1.0     &       0.92    &       CAS     \\
2\      454\    953.769 &       75      &       $-208$  &       1.0     &       0.13    &       ESO     \\
2\      454\    955.765 &       $-82$   &       256     &       1.0     &       0.88    &       ESO     \\
2\      454\    956.780 &       95      &       $-272$  &       1.0     &       0.26    &       ESO     \\
2\      454\    960.790 &       -90     &       280     &       1.0     &       0.76    &       LCO     \\
2\      454\    961.818 &       68      &       $-218$  &       1.0     &       0.15    &       LCO     \\
2\      454\    961.830 &       71      &       $-226$  &       1.0     &       0.15    &       LCO     \\
2\      454\    963.715 &       $-100$  &       328     &       1.0     &       0.86    &       LCO     \\
2\      454\    963.723 &       $-95$   &       308     &       1.0     &       0.86    &       LCO     \\
2\      454\    963.921 &       $-69$   &       247     &       1.0     &       0.93    &       LCO     \\
2\      454\    964.662 &       92      &       $-261$  &       1.0     &       0.21    &       LCO     \\
2\      454\    964.922 &       95      &       $-265$  &       1.0     &       0.31    &       LCO     \\
2\      454\    984.868 &       $-90$   &       310     &       1.0     &       0.79    &       CAS     \\
2\      455\    846.507 &       $-88$   &       283     &       1.0     &       0.78    &       LCO     \\
2\      456\    521.470 &       $-89$   &       250     &       1.0     &       0.80    &       CAS     \\
2\      456\    521.678 &       $-82$   &       246     &       1.0     &       0.88    &       CAS     \\
2\      456\    522.473 &       86      &       $-219$  &       1.0     &       0.18    &       CAS     \\
2\      456\    522.659 &       102     &       $-255$  &       1.0     &       0.25    &       CAS     \\
2\      456\    523.605 &       $-29$   &       67      &       1.0     &       0.60    &       CAS     \\
2\      456\    523.707 &       $-41$   &       127     &       1.0     &       0.64    &       CAS     \\
2\      456\    524.524 &       $-67$   &       193     &       1.0     &       0.95    &       CAS     \\
2\      456\    524.687 &       $-31$   &       71      &       1.0     &       0.01    &       CAS     \\
\noalign{\smallskip}\hline\noalign{\smallskip}
\end{tabular}
\end{table}

\subsection{Separated spectra}

\begin{figure*}
\includegraphics[angle=270,scale=0.70]{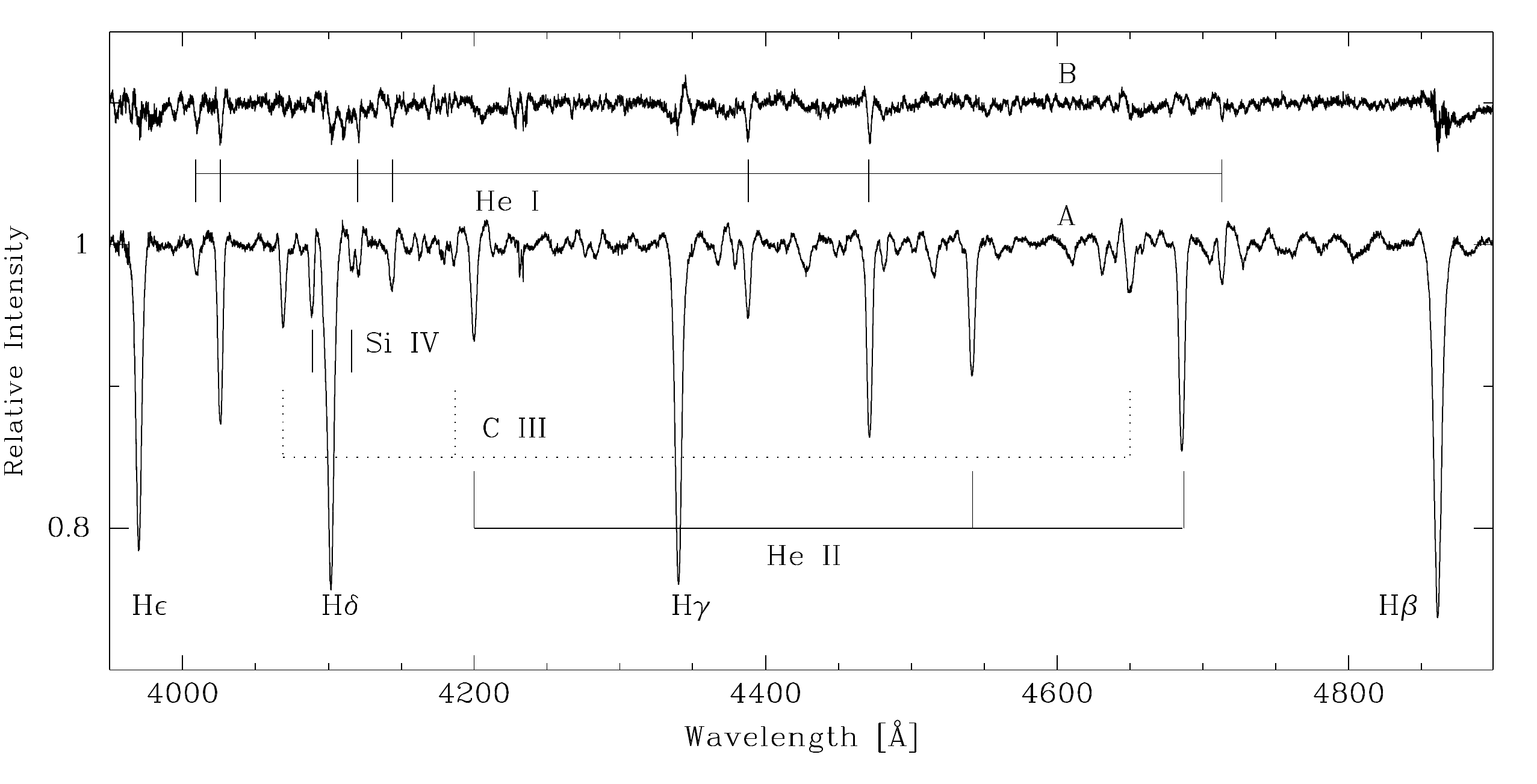}
\caption{Separated optical spectra of HD~161853 A and B. The most conspicuous lines are 
labelled.}
\label{spec}
\end{figure*}

We independently analysed and classified the separated spectra following
\citet{2011ApJS..193...24S} for the O-type primary, and following
\citet{1990PASP..102..379W} for the 
B-type secondary component.
We inspected the spectra visually using the {\sc mgb} code \citep{2012ASPC..465..484M}, which allows the
user to compare the unknown spectrum with standard stars of each sub-type.

We determined a spectral type O8~Vz for the primary and B1--3~V for the secondary. The O8~Vz 
classification agrees with \citet{gosss2}.
The B1--3 sub-type is only determined from the intensity relations among the faint He~{\sc i} lines. Other
lines useful as primary criteria, such as
Mg {\sc ii} $\lambda4481$\,\AA\ and Si~{\sc ii} $\lambda\lambda4128,4130$\,\AA, are very noisy or affected by
residuals in the derived spectrum.
The two spectra are shown in Fig.~\ref{spec}.

\subsection{Orbital solution}

The obtained RV measurements of the primary star were used to search for periodicities by means of the
Lomb-Scargle algorithm \citep{1982ApJ...263..835S}. This algorithm is 
provided on-line as a NASA Exoplanet Archive service
\citep{2013PASP..125..989A}. A period of 2.66 days was obtained.

The orbital solution of the SB2 was determined with an
improved version of the original program for the determination of the orbital elements of
spectroscopic binaries \citep{1969RA......8....1B}, named {\sc gbart} and
developed by F. Bareilles\footnote{{\sc gbart} is available at 
http://www.iar.unlp.edu.ar/\textasciitilde fede/pub/gbart.}.
Some RV values were weighted by 0.5 when the two components did not separate well in the respective
spectra (see Table~\ref{vrs}).
The RVs of both components converged to a slightly eccentric orbit and a relatively low mass ratio. 
The orbital elements are shown in Table~\ref{parameters} and the RV curves are depicted in Fig.~\ref{vrs}.

\begin{table}
\begin{center}
\caption{Orbital parameters of HD~161853\,AB.} \label{parameters}
\begin{tabular}{lc}
\hline\hline\noalign{\smallskip}
Parameter & value \\
\noalign{\smallskip}\hline\noalign{\smallskip}
$P$ [d]                              &    2.66765$\pm$0.00001\\
$T_{\rm peri}$ [HJD]                 & 2\,456\,634.04$\pm$0.02   \\
$T_{\rm RV\,max}$ [HJD]              & 2\,456\,634.73$\pm$0.02   \\
$V_0$ [km s$^{-1}$]                  &        4.0$\pm$1.0    \\
$e$                                  &      0.121$\pm$0.007  \\
$\omega$ [deg]                       &        254$\pm$4      \\
$q$($M_{2}$/$M_{1}$)                 &      0.332$\pm$0.007  \\
$K_{1}$ [km s$^{-1}$]                &         96$\pm$2      \\
$K_{2}$ [km s$^{-1}$]                &        287$\pm$2      \\
$M_{1}~\sin^{3} i$  [M$_{\odot}$]    &       11.4$\pm$0.4    \\
$M_{2}~\sin^{3} i$   [M$_{\odot}$]   &        3.8$\pm$0.3    \\
$a_{1}~\sin i$ [R$_{\odot}$]         &       4.97$\pm$0.08   \\
$a_{2}~\sin i$  [R$_{\odot}$]        &      14.95$\pm$0.09   \\
\noalign{\smallskip}\hline
\end{tabular}
\end{center}
\end{table}

\begin{figure}
\includegraphics[scale=.73]{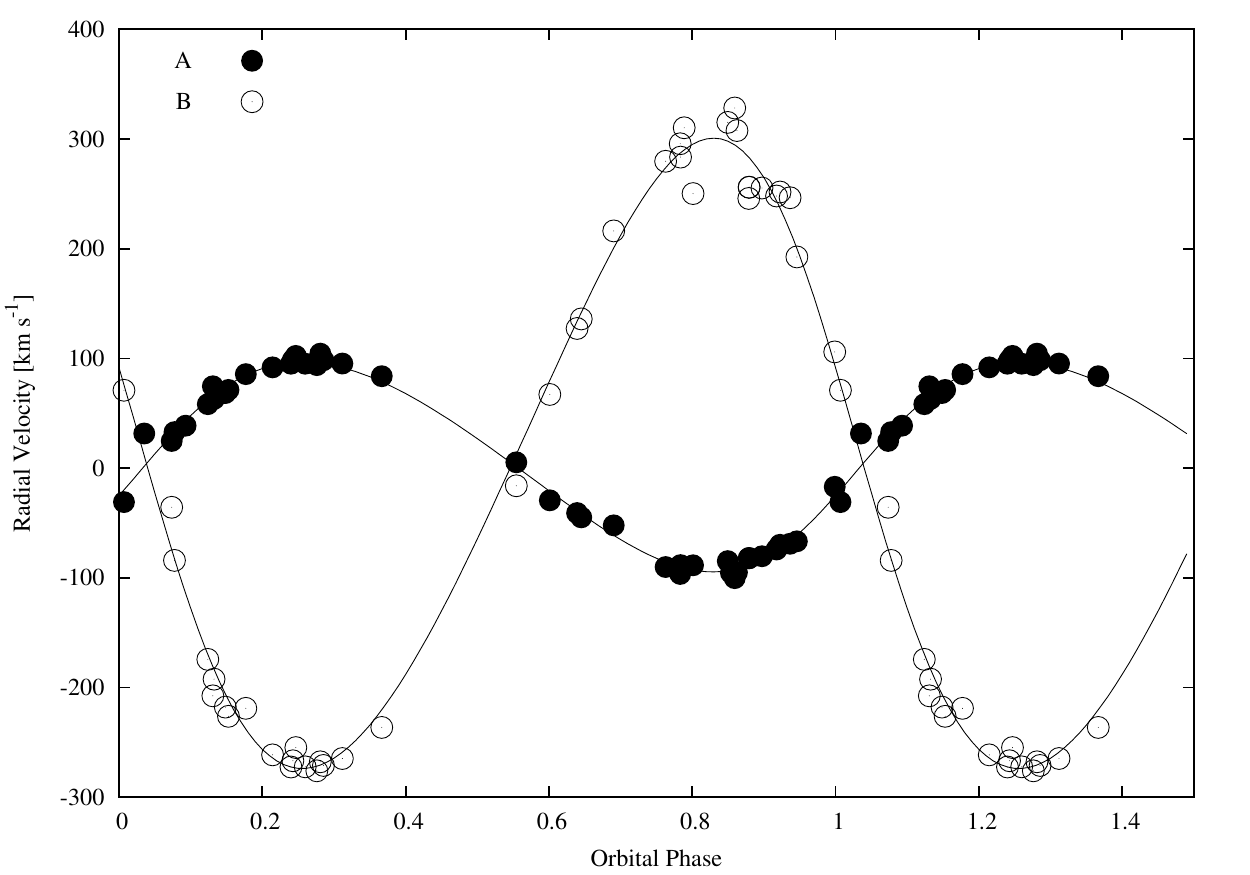}
\label{sb2}
\caption{
RV curves of the two components in HD~161853. RV errors are smaller than the size of the symbols.
The curves depict the orbital motion calculated from the parameters of Table~\ref{parameters}.
}
\end{figure}

The non-zero orbital eccentricity implies that the age of the system should be younger than the
expected circularisation time, $t_{circ}$. This value was found to be $t_{circ} \sim 1$ Ma
using Eq. 5.9 in \citet{1977A&A....57..383Z} and adopting the mass and radius from
\citet{2005A&A...436.1049M}
and the tidal torque constant $E_2$=3.5~$10^{-6}$ \citep{1975A&A....41..329Z}.
This means that HD~161853 is a young massive system.

Through comparison of the minimum masses in Table~\ref{parameters} with the masses expected from 
theoretical models shown in Table~\ref{teoria}, we estimated possible inclinations of 
$i$=54$\pm$7~deg (using the O star mass) and $i$=47$\pm$7~deg (using the B star mass range). 
The theoretical mass ratio obtained adopting a B3~V sub-type for the secondary star agress better with the observations than adopting B1 V.

\begin{table}
\begin{center}
\caption{Theoretical stellar parameters.} \label{teoria}
\begin{tabular}{r c c c}
\hline\hline\noalign{\smallskip}
Parameter & O8 V & B1 V & B3 V \\
          &  Martins et al. (2005)    & \multicolumn{2}{c}{Cox (2000)}\\
\noalign{\smallskip}\hline\noalign{\smallskip}

$M$ [M$_{\odot}$]   & 21.5 & 14.2 & 7.6  \\
$R$ [R$_{\odot}$]   & 8    & 6.5  & 4.8   \\

\noalign{\smallskip}\hline
\end{tabular}
\end{center}
\end{table}

\subsection{Photometric analysis}

The short period of the binary encouraged us to retrieve and analyse the photometry obtained by the 
All Sky Automated Survey \citep[ASAS,][]{2002AcA....52..397P}.
The data, although near the saturation limit, appear to be almost constant between 2001 and 2009. 
Statistics performed over 813 values (six-pixel aperture photometry with quality labelled as A) 
gave a mean of $V$=7.956 mag, a standard
deviation of 0.014 mag, and a range of 0.084 mag between the lowest and highest values. Thus, the
lack of eclipses
can also be used to constrain the highest orbital inclination.
We fitted the spectroscopic and photometric data with a Wilson-Devinney
model \citep{1971ApJ...166..605W} by means of the {\sc phoebe} code \citep{2005ApJ...628..426P}
and adopting the theoretical stellar parameters shown in Table~\ref{teoria}.
We determined a highest orbital inclination of 
$i$=54~deg, and hence the masses should be greater than 22 M$_\odot$ and 7.2 M$_\odot$ for the
O8~V and B1--3~V stars, respectively.
We note the agreement between 
the value of the inclination angle from photometry and the one
from the comparison between the minimum masses and those expected from the theoretical models.

The available multi-band photometry permits analysing the
interstellar extinction and determination of the spectrophotometric distance to HD~161853. We
applied the {\sc chorizos} code \citep{chorizos} to the $UBV$ \citep{2008RMxAC..33...56S},
Tycho-2 $BV$ \citep{2000A&A...355L..27H}, and $JHKs$ \citep[2MASS; ][]{2006AJ....131.1163S} data.
We obtained 
$E(4405-5495)=0.496\pm0.011$~mag,
and $A_{\rm V}=1.82\pm0.02$~mag, which results in a distance modulus $V_{J,0}-M_V$=6.242-$M_V$~mag. 
Adopting $M_V=-4.3$~mag for the O8~Vz star \citep{1972AJ.....77..312W,2005A&A...436.1049M} and
that the effect of the flux of the B1--3~V star is smaller than the uncertainty (0.45~mag), we derived a 
distance of $1.35\pm0.25$~kpc, which is consistent with the kinematic distance to the CO molecular cloud 
\citep{2014A&A...569A.125H} and to the Dutra-Bica~51 open cluster \citep{2002A&A...389..871D}.
This determination supports the HD~161853 membership in a star-forming region that should be
studied with further tailored observations.

\begin{figure*}
\includegraphics[scale=.48]{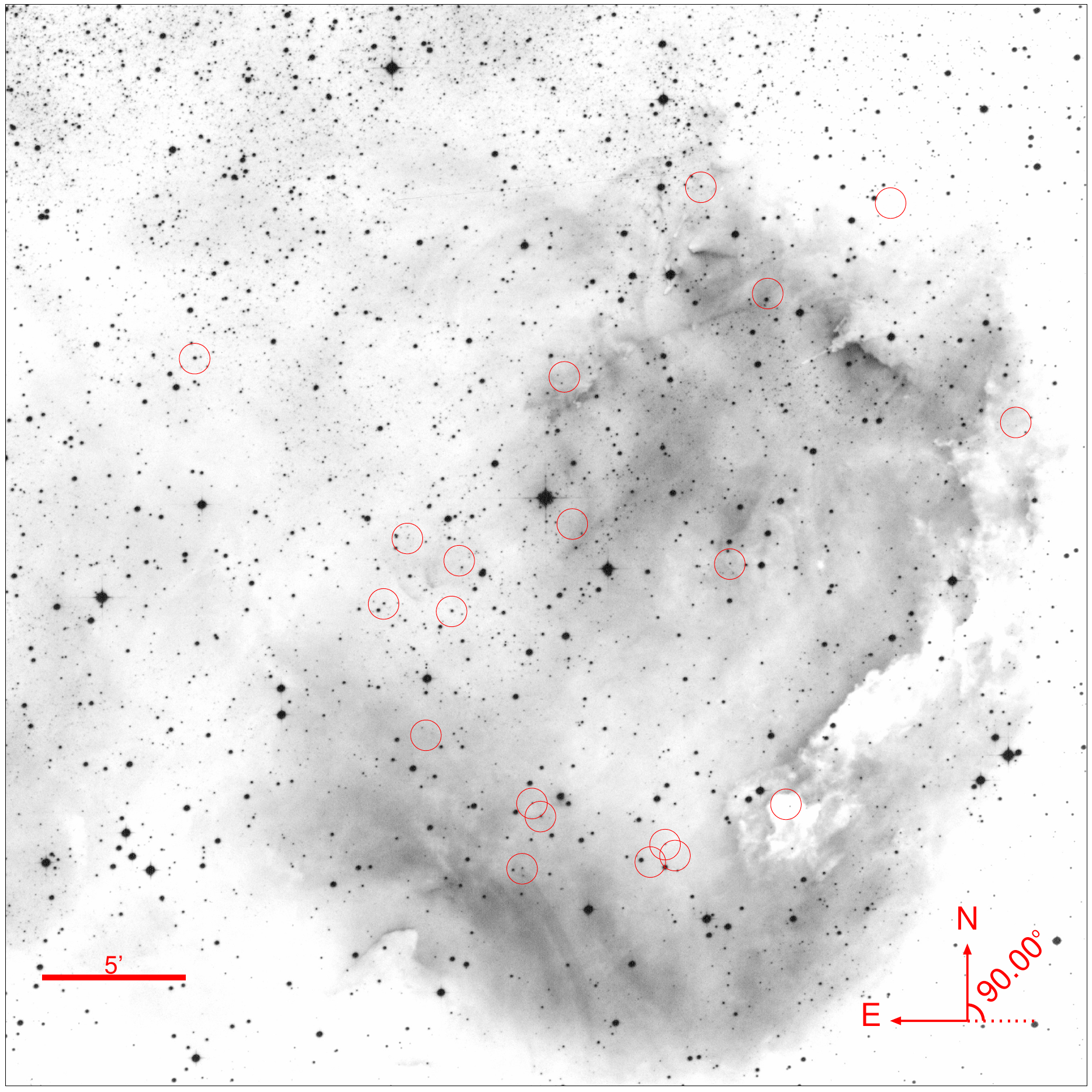}
\includegraphics[scale=.48]{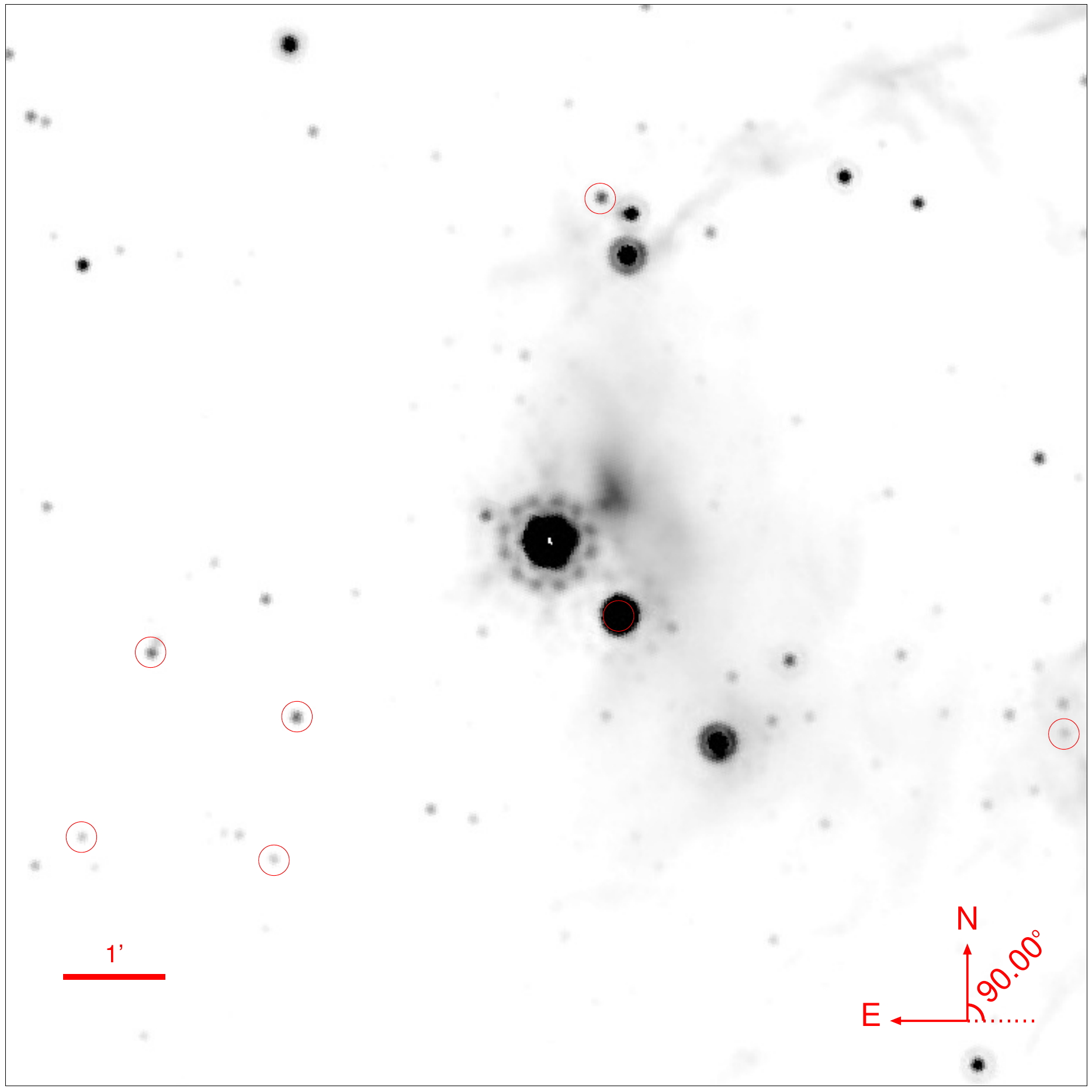}
\caption{Images centred on HD~161853. Young stars are indicated with circles.
{\it Left panel:} $30\times30$ arcmin SuperCOSMOS AAO/UKST H$\alpha$ image
\citep{2005MNRAS.362..689P} showing the Sh~2-15 H {\sc ii} region.
{\it Right panel:} $5.5\times5.3$ arcmin {\it Spitzer} 24 $\mu$m image showing some structures associated with 
our star. 
}
\label{spitzer24mu}
\end{figure*}

The mid-infrared images of {\it WISE} \citep[22 $\mu$m;][]{2010AJ....140.1868W} and 
{\it Spitzer/MIPS} \citep[24 $\mu$m;][]{2004ApJS..154...25R}
reveal a compact emission region 50 arcsec north-west from HD~161853 (see Fig.~\ref{spitzer24mu}).
This feature is not detected in the shorter wavelength bands of {\it Spitzer} and {\it WISE}.
It may be a dust wave, similar to the arc-like cloud near $\sigma$~Ori
noted by \citet{2008A&A...491..515C}. \citet{2014A&A...563A..65O} explained these regions as the 
result of surrounding dust piled up by the radiation pressure of the massive star.

\section{Summary}

The O-type star HD~161853 is an SB2 system with massive components. We separated the
composite spectrum with a separation method and classified the primary star as O8~Vz
and the secondary as B1--3~V. We determined individual RVs for both components and
derived the orbital parameters, obtaining a period of 2.66765 days and an eccentricity of 0.121.
We also calculated minimum masses of 11.4 M$_\odot$ and 3.8 M$_\odot$ for the O- and B-type stars,
respectively.

We analysed the photometry of this star available in the ASAS database.
The photometry was useful to constrain the orbital inclination to
a value lower than 54~deg and hence increasing the minimum masses to 22 and 7.2 M$_\odot$.

The minimum masses determined for the stars and the eccentricity are the most direct and reliable
proof that the O-type component is a massive young star and not a post-AGB object.  
It is located in a very young region in the centre of an H {\sc ii} region, 
consistently with the results of the binary analysis.
We have identified a compact emission region 50 arcsec north-west from our star. The mid-infrared 
flux from this dust wave caused HD~161853 to be erroneously considered as a post-AGB star for almost
two decades.

\begin{acknowledgements}
We thank the referee Jos\'e A. Caballero for careful reading of our paper and useful suggestions that 
improved the work.
We thank the directors and staffs of CASLEO, LCO, and ESO/La Silla for the use of their facilities
and their kind hospitality during the observing runs.
CASLEO is operated under agreement between CONICET and the
Universities of La Plata, C\'ordoba and San Juan, Argentina.
The \'Echelle Li\`ege Spectrograph was jointly built by REOSC and Li\`ege Observatory and is on long-term loan from the latter.
RHB acknowledges FONDECYT Project No. 1140076.
JIA aknowledges financial support from FONDECYT Project No. 11121550.
This research has made use of NASA's Astrophysics Data System, the SIMBAD database operated at CDS, Strasbourg, France, and 
the Aladin interactive sky atlas developed at CDS.
\end{acknowledgements}

\bibliographystyle{aa}
\bibliography{own}
\end{document}